\documentclass[aps, prb, twocolumn,floatfix, showpacs, 10pt]{revtex4}
\usepackage{amsmath}
\usepackage{amsfonts}
\usepackage{amsbsy}
\usepackage{amssymb}
\usepackage{graphicx}
\usepackage{textcomp}
\usepackage[caption=false,singlelinecheck=false]{subfig}
\usepackage{xcolor}
\usepackage{mathrsfs}
\usepackage{bm}
\usepackage{mathrsfs}
\usepackage{ulem}
\usepackage[colorlinks,linkcolor=red,citecolor=blue,filecolor=magenta,
urlcolor=cyan]{hyperref}

\allowdisplaybreaks

\begin{document}
\title{Topological multiferroic phases in the extended Kane-Mele-Hubbard Model in the Hofstadter regime}

\author{Archana Mishra}

\author{SungBin Lee}

\affiliation{Korea Advanced Institute of Science and  Technology, Daejeon, South Korea}

\date{\today}
\begin{abstract}
{We investigate the new quantum phases on the extended Kane-Mele-Hubbard model of honeycomb lattice in the Hofstadter regime. In this regime, orbital motion of the electrons can induce various topological phases with spontaneously broken symmetries when the spin orbit coupling and electron correlations coexist. Here, we consider the interaction effects in the Kane-Mele model and discuss possible phases in the presence of magnetic field at integer fillings of electrons. In particular, focusing on $2\pi/3$ magnetic flux per plaquette, the realization of numerous quantum phases are discussed within the mean field framework;
insulator with coplanar magnetic ordering, ferrimagnetic Chern insulator with nematic charge order, ferrimagnetic-ferrielectric Chern insulators etc.
Many of these phase transitions are also accompanied with the change in the topological invariants of the system. Based on our theoretical study, we propose topological {\it multiferroic} phases with a scope of realization in 2D van-der Waals materials and optical lattice system where the significant interplay of magnetic field, spin orbit coupling and interactions can be engineered.}
\end{abstract}
\maketitle
%
%

\section{\label{sec:I}Introduction}
Initiated by the discovery of graphene, the two dimensional van-der Waals materials and their interesting properties have gotten a lot of attention in condensed matter physics. 
In particular, the successful synthesis of 2D van-der Waals materials with various distinct ions and their multilayers has enhanced the scope of the realization of new quantum phases. Recent studies have explored and observed novel strong electron correlation effect and unconventional superconductivity in the gate tuned multilayer systems \cite{volovik1991,cao2018,chen2018,xu2018,po2018,roy2018,dodaro2018}.
 In addition, couple of transition metal chalcogenides series are recently reported as the magnetic materials exhibiting magnetic orderings either ferro-type or antiferro-type where the interplay of the spin-orbit coupling and electron correlations are crucial to understand the physical properties of these systems \cite{kim2018,gong2017,chittari2016,sivadas2015,lin2016,lee2016,park2016,sugita2017,mishra2018}.

Along with the discovery of two dimensional materials, the strong magnetic field effect also has been investigated in the lattice system. In particular, the realization of Moir\'{e} pattern by a single layer graphene on top of a boron-nitride crystal allowed a periodicity much larger than the lattice constant, thus, enabling the experimental realization of the fractal quantum Hall effect of the Hofstadter spectrum which had long been a theoretical concept \cite{ponomarenko2013,dean2013,hunt2013,yu2014,hofstadter1976}.
When both strong magnetic field and spin orbit coupling (SOC) coexist, their combination could give rise to interesting topological phases. As one example, the topological phase transitions in the non-interacting Kane-Mele model of honeycomb lattice have been studied as a function of magnetic flux and  SOC strength for different electron fillings \cite{beugeling2012, goldman2012}. These topologically non-trivial phases have distinct Chern numbers for each spin and are characterized by the Hall conductivities of the system which change with the band gap closing.
Motivated by those topological phases, one could further ask what is the role of electron correlations. In the presence of strong electron interactions, one may expect correlation driven phase transitions in topologically non trivial phases which we study in this paper. 

Many kinds of Landau type phase transitions can accompany with the topological phase transitions as a consequence of electron interactions. Interactions can induce spontaneous symmetry breaking of the lattice giving rise to numerous exotic phases. There are many studies related to the interaction driven quantum Hall and quantum spin Hall phases \cite{raghu2008,weeks2010,wen2010,grushin2013,araujo2013,castro2013}. 
In addition, the effect of interactions in the Hofstadter regime
has also been studied extensively, showing spontaneous breaking of translational and rotational symmetries of the system prompting
topological transitions\cite{gudmundsson1995,doh1998,czajka2006,chakraborty2013,apalkov2014,archana2016,archana2017}. Similarly, interplay of interactions and SOC on the lattice system has also been widely explored and several exotic phases such as Mott topological insulators and magnetic Chern insulators have been investigated as an outcome
\cite{chung2014,budich2012,rachel2010,wu2012,lang2013,araujo2013,pesin2010,jackeli2009,raghu2008,mishra2018}. 

All these interesting phases are quite challenging to find in real materials. In particular, the apt area to experimentally realize the combined effect of magnetic field, SOC and strong electron correlations are artificial lattice systems. For instance, the tunable Moir\'{e} pattern in graphene superlattices can realize the Hofstadter regime in addition to strong correlation effect \cite{yang2016, kim2017}. Thus by gating those graphene superlattices, one can realize different electron fillings and expect correlation driven topological phases. 
The superlattice structures of transition metal trichalcogenides are also good candidates having considerable SOC and electron correlations. 
Another prime area where this system can be realizable is the optical lattice. Both uniform and staggered magnetic fluxes per plaquette have been synthesized in 
ultracold atomic system \cite{aidelsburger2011,aidelsburger2013a, aidelsburger2013, miyake2013}. 
Here, the complex hoppings originate from the synthetic gauge fields
induced through laser assisted
tunneling or periodic optical lattice shaking  \cite{jaksch2003,gerbier2010,aidelsburger2011,aidelsburger2013,aidelsburger2013a,struck2012,jotzu2014,goldman2014}. The effect of SOC can also be
introduced into the optical lattice using similar methods and has been explored in many contexts \cite{goldman2010,cooper2011,hauke2012,mazza2012,kennedy2013,beeler2013,struck2014}.
Hence, taking into account these possible realization in both superlattices of two dimensional van-der Waals materials and ultracold atomic systems, we theoretically propose new quantum phases where SOC, magnetic field and electron correlations all play a significant role together.

In this paper, we study the electron correlation effect when both considerable SOC and magnetic field coexist and discuss a class of topological phases where the Landau type order parameters also become  finite driven by strong correlations. As a minimal model, we explore the Hofstadter-Harper Hamiltonian for the honeycomb lattice focusing on a special flux ($\phi=2\pi/3$), in the presence of intrinsic SOC and onsite and nearest neighbor interactions\cite{harper1955,hofstadter1976}. One of the intriguing results is the discovery of topological {\it multiferroic} phases where in addition to non zero Chern numbers, both nematic/ ferrielectric orderings and various types of magnetic orderings are simultaneously stabilized. On symmetry grounds, interactions give rise to the phases which break not only the inversion symmetry but also the translational symmetry and rotational symmetry of the system. Beyond the conventional charge order accompanied
with ferro-, ferri- or antiferro-magetic types of orderings, the breaking of translational and rotational symmetries can lead to incompressible nematic and ferrielectric phases characterized by the dipole and quadrupole moments\cite{archana2016}. In addition, the presence of SOC and magnetic flux further induce the Chern insulating phases accompanied with these symmetry broken magnetic and electric phases. We note that such Chern insulators with Landau type order parameters can also naturally induce not only the change in the Hall conductivity but also the staggered magnetic flux and non zero orbital currents. 

This paper is organized as following: Section II describes the minimal model and its symmetry arguments. Here, we also briefly review the band structure and its topological nature for non-interacting system. In Section III, the interacting Hamiltonian and the mean field technique to solve this interacting problem is discussed. The results are outlined in Section IV and we summarize the paper in Section V.

\section{\label{sec:II} Model and  its symmetries}
We consider the Kane-Mele model in the Hofstadter regime in the presence of onsite and nearest neighbor interactions. When the strong magnetic field is applied, one can expect two main effects; orbital motion of electrons and a Zeeman splitting. In our study, we focus on the effect of orbital motion of electrons without including a Zeeman coupling term. With a dominant Zeeman coupling, electron bands with spin up and bands with spin down are completely split, thus, one can argue electron interaction effect with fully spin polarized case which is rather trivial.

The model we study is described by the following Hamiltonian,
\begin{align} 
\label{eq:1} 
H=&-t\sum_{\langle ij\rangle\sigma}e^{i A_{ij}}\left(c^\dagger_{i\sigma}c_{j\sigma}+h.c\right)\nonumber\\
&-i\lambda_{so}\sum_{\langle\langle ij\rangle\rangle, \sigma}
e^{i A'_{ij}}\left(c^\dagger_{i\sigma} \nu_{ij}s^z c_{j\sigma}+h.c\right)\nonumber\\
&+U\sum_{i}n_{i\uparrow}n_{i\downarrow}+V\sum_{\langle  ij\rangle}n_{i}n_j,
\end{align}
where $c_{i\sigma}~(c_{i\sigma}^{\dag})$ is the annihilation (creation) operator for
electrons at site $i$ on the honeycomb lattice with spin $\sigma$, $n_{i\sigma}\!=\!c^\dagger_{i \sigma} c_{i \sigma}$ is the number density
operator at spin $\sigma \in (\uparrow, \downarrow)$, $n_{i}\!=\! n_{i \uparrow} + n_{i \downarrow}$,
$t$ is the nearest  neighbor hopping parameter, $\lambda_{so}$ is the strength of intrinsic SOC between the next nearest neighbors, 
$U$ is the onsite interaction strength and $V$ is the nearest  neighbor interaction strength. 
$s^z$ represents the $z$ components of the Pauli matrices for spin and
$A_{ij}$, $A'_{ij}$ are gauge potentials on  the nearest neighbor  and next-nearest  neighbor links such that the
flux per plaquette is  $2\pi/q$  where  $q$ is an integer. $\langle ij\rangle$ denotes nearest neighbor
and $\langle\langle ij\rangle\rangle$ denotes next nearest neighbor between site $i$ and site $j$. The sign $\nu_{ij}=\pm1$ depends on the
value of the outer product of two bond vectors connecting site $i$ and $j$ through a common neighboring site $l$. From now on, we set $t=1$ and $\lambda_{so}$, $U$  and $V$ are in units of $t$. 

The Hamiltonian is invariant  under three fold and two fold rotations. 
In the presence of magnetic flux $2\pi/q$ per plaquette, the Hamiltonian commutes with the two magnetic translation operators $\tau_1$ and  $\tau_2$ which are along the basis vectors of the original honeycomb lattice.
$\tau_1$ and $\tau_2$  don't commute with each other, $\tau_1\tau_2\tau_1^{-1}\tau_2^{-1}=e^{i\frac{2\pi}{q}}$.
Hence, in order to implement the Bloch theory, the magnetic unit  cell has to be $q$ times larger than the original unit cell.
Here, we consider the case  of
$q\!=\!3$ i.e. $2\pi/3$ flux per plaquette and understand how the combination of electron orbital motion and SOC could induce various topological phases in the presence of interactions. Although our results are focusing on a particular value of magnetic flux, the same analysis can be applied for different fluxes and expect similar  electric and magnetic phases with non trivial topology. 
Before we discuss electron correlation effect, let's briefly review the Kane-Mele model in the Hofstadter regime for the  non-interacting case. 

\subsection{Review : Non-interacting case ($U=V=0$)}

In this section, we summarize the results of the Kane-Mele model in the Hofstadter regime when the electron-electron interactions are absent. 
In this limit, possible phases has been explicitly studied by Beugeling et. al\cite{beugeling2012, goldman2012}. They investigate a variety of 
two dimensional topological phases that arise from the competition of intrinsic spin orbit coupling, Rashba spin orbit coupling, Zeeman splitting in the presence of uniform magnetic field. While the SOC results in helical spin current at edge and realizes quantum spin Hall (QSH) phases, magnetic field induces chiral  spin current at edges resulting in time reversal symmetry broken quantum Hall (QH) phases.
Thus, the presence of both SOC and magnetic field in the system can induce the topological  phase transition between helical QSH phase and chiral spin imbalanced QH phase. 

\begin{figure}[hbtp]
\begin{center}
\includegraphics[scale=0.15,trim=12cm 0mm 2cm 0.2cm,clip]{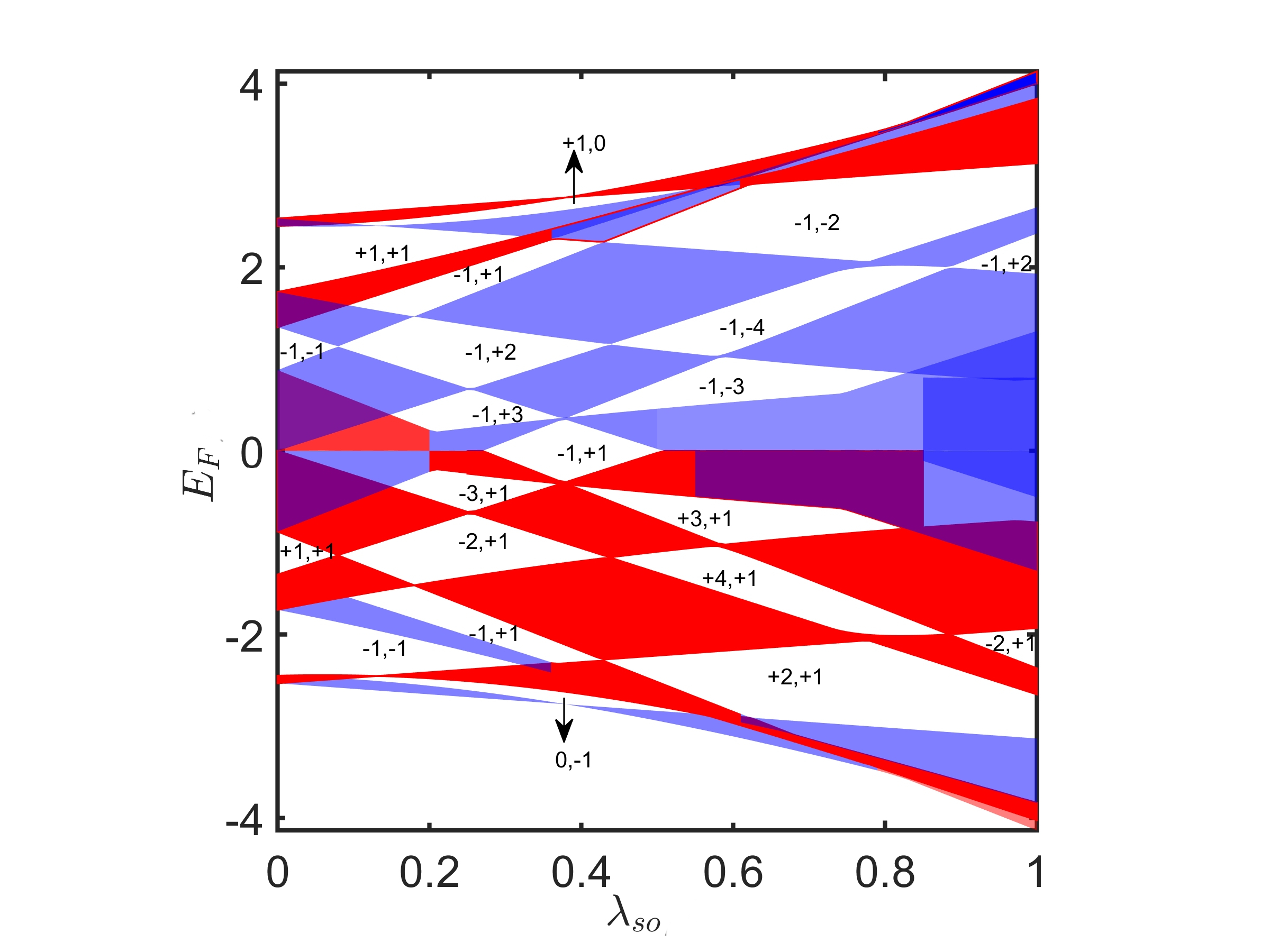}\\
\caption{\label{fig0}(Color online) Non-interacting case : Phase diagram as functions of Fermi energy ($E_F$) and intrinsic spin orbit coupling ($\lambda_{so}$) for flux  per plaquette $2\pi/3$.  
In the colored region, the system is metallic. The colors red and blue distinguish the energy dispersion for spin up and down bands respectively. The numbers in the phase diagram indicate the total sum of Chern numbers for occupied bands of each spin component; left for spin up bands, right for spin down bands.}
\end{center}
\end{figure}

As mentioned earlier, the magnetic flux $\phi \!=\! 2\pi/3$ triples the original unit cell of honeycomb lattice and results in total 12 distinct energy bands in the momentum space (3 for a magnetic unit cell $\times$ 2 for sublattices $\times$ 2 for spins $\uparrow$,$\downarrow$). The particle-hole symmetry connects the lower half and the upper half bands, which also guarantees the zero Hall conductivity at half filling if the band gap exists. Similarly, the Hall conductivity for upper half bands are related to the opposite signs of the Hall conductivity for the lower half bands.
Fig.\ref{fig0} is the phase diagram as functions of the Fermi energy $E_F$ and the strength of intrinsic SOC, $\lambda_{so}$, in the presence of magnetic flux $\phi \!=\! 2\pi/3$, which has also been studied in Ref.~\onlinecite{beugeling2012}. 
 The red and blue colors distinguish the energy bands for spin up and down components respectively. The  Chern  numbers are indicated for spin up and spin down components where the gap exists in the spectrum; The left (right) number corresponds to the Chern numbers summed over all the occupied spin up (down) bands. 

As shown in Fig.\ref{fig0}, there are many topological transitions  seen due to the SOC.  
For instance, at filling $\nu=6$ ($\nu$ is the number of electrons per magnetic unit cell), there is a region for $\lambda_{so}=[0.25,0.5]$ where the time reversal symmetry broken QSH phase is stabilized. 
For other integer fillings i.e. integer $\nu$, there are some interesting
phases such as spin  filtered  phases and spin imbalanced phases that arise because of the interplay of SOC  and magnetic field.
In the spin  filtered phase, there is a chiral edge  current where the contribution is only from a single spin  component. In case of spin imbalanced  phase, there is a helical spin current at 
the edge where both spin up and down components contribute but they are not equal resulting in a  spin imbalance phase which is distinct from the normal QSH  phase.
Based on the phase diagram shown in Fig.\ref{fig0}, we study the electron interaction effect and investigate  how these  topological phases and  phase transitions are affected. 

\section{Effect of interactions ($U\neq 0$, $V\neq0$)}

When electron interactions are present, one could expect the Landau type phase transitions by stabilizing charge or spin order parameters. In particular, the strong onsite Coulomb interaction could induce magnetic order, while the  off site Coulomb interaction could induce charge order in addition to the modification of electron hoppings of the system. Apart from the conventional charge and magnetic order that break the inversion symmetry of the system, these interactions can lead to breaking of additional lattice symmetries resulting in some exotic phases. Furthermore, one could expect the coexistence of magnetic field and SOC, in the presence of interactions, giving rise to these Landau type phase transitions accompanied with the change in the topological invariants of the system. To explore these aspects mentioned above, we consider onsite interaction $U$ and nearest neighbor interaction $V$ for the minimal model (See Eq.\eqref{eq:1}) and investigate possible topological phases and their phase transitions based on the Hartree-Fock mean field analysis. Due to the absence of the Fermi surface nesting at each integer filling, the spin, charge and bond order parameters $\bm{M}$, $\Delta_i$, $\chi_{ij}^{\sigma \sigma'}$ are introduced within the magnetic unit cell.
The  mean field Hamiltonian is given as,
\begin{eqnarray}
\label{eq:4}
\mathcal{H}_{mf}&=&\mathcal{H}_0
\!+\! U \sum_{i}  (\frac{\Delta_i}{2} {n_i} -  \bm{M}_i \cdot \bm{S}_i) 
\nonumber \\
&&\!+V \sum_{\langle ij \rangle} \Big( (\Delta_i n_j  + \Delta_j n_i)- \sum_{\sigma \sigma'} (\chi_{ij}^{\sigma \sigma'}c_{j \sigma'}^\dagger c_{i\sigma} +h.c) \Big) \nonumber \\
&& \!- \frac{U}{4} \sum_i \Big( {\Delta_i^2} \!-\! {\bm{M}_i^2}  \Big)\!-\! V \sum_{\langle i j \rangle} \Big( \Delta_i \Delta_j \!-\! \sum_{\sigma \sigma'} |\chi_{i j}^{\sigma \sigma'}|^2 \Big)\nonumber\\
\end{eqnarray}
where $\mathcal{H}_0$ represents the non-interacting Hamiltonian in Eq.~\eqref{eq:1}. The self consistency equations for spin, charge and bond order parameters are,
\begin{eqnarray}
\label{eq:5}
 \bm{M}_i&=&2\langle\bm{{S}}_i\rangle \nonumber\\
 \Delta_{i}&=&\langle n_i\rangle \nonumber\\
 \chi_{ij}^{\sigma \sigma'}&=&\langle c^\dagger_{i\sigma}c_{j\sigma'}\rangle,
\end{eqnarray}
with the spin operator $\bm{S}_i= \frac{1}{2}c^\dagger_{i \sigma}{\bm{s}_{\sigma \sigma'}}c_{i \sigma'}$ ($\sigma, \sigma' \in \{\uparrow, \downarrow\})$ with the Pauli matrices $\bm{s}$ and the number density operator at site $i$ is $n_i\!=\!n_{i \uparrow}\!+\!n_{i \downarrow}$.
Then, the mean field Hamiltonian matrix is written by $12\! \times \!12$  matrix in momentum space, $h_{mf}(\bm k)=h_0(\bm k)+ h_1(\bm k,\bm M, \Delta,\chi)$ where $h_0(\bm k)$ is the Hamiltonian matrix for $\mathcal{H}_0$ 
and $h_1(\bm k,\bm M, \Delta,\chi)$ is the Hamiltonian matrix for interaction part after the mean field decoupling.
In terms of order parameters, there are in total 6 magnetization vector order parameters ($\bm M_{\alpha,a}$),
6 real charge order parameters ($\Delta_{\alpha,a}$) and 36 complex order bond parameters $\chi_{(\alpha,a)(\beta,b)}^{\sigma \sigma'}$.
Here, the sites $i$ and $j$ are rewritten with the labels $(\alpha,a)$ and $(\beta,b)$ respectively where the indices $\alpha,\beta \! \in \! \{1,2,3\}$ are for the tripled  magnetic unit cell in the presence of magnetic flux $2\pi/3$ per plaquette and
$a,b \! \in \! \{A,B\}$ denote the sublattices. 

The self consistency equations in Eq.~\eqref{eq:5} are solved 
for the values of $\lambda_{so} \! \in \![0,1]$, $U \! \in \![0,8]$ and $V \! \in \![0,8]$. The number density  is fixed such that the self consistency equations are solved at each integer filling $\nu \!\in\![1,12]$. To investigate all possible phases at every integer filling, one needs to consider $\nu \!\in\![1,6]$ in the presence of particle-hole symmetry of the system.


\section{Results}
In this  section, we  discuss  the  topological   phases realized at each integer filling, $\nu$, based on the mean field analysis mentioned above. For finite but small $U$ and $V$, the phases for non-interacting case are still stable at all fillings.
We denote this as the symmetric phase ($\mathcal{S}$). The charge density is uniform without developing any magnetization and and all the
symmetries of the Hamiltonian are preserved in these phases. With increasing $U$ and $V$, we find several
charge and magnetic ordered phases that break translational, rotational and inversion symmetries of the system. To investigate specific electric properties induced by charge order, we characterize the dipole moment
($P^\mu$) and the quadrupole moment ($Q^{\mu\nu}$) defined with the charge order parameter $\Delta_i$ as following\cite{archana2016},
\begin{align}
&P^\mu\equiv\frac{1}{N_{tot}}\sum_i~R^\mu_i\Delta_i,\\
&Q^{\mu\nu}\equiv\frac{1}{N_{tot}}\sum_i\left(2R^{\mu}_iR^\nu_i
-\delta^{\mu\nu}\bm{R_i}\cdot\bm{R_i}\right)\Delta_i~,
\end{align}
where $R^\mu_i~(\mu=1,2)$ are the components of the position vector $\bm{R_i}$
at site $i$, and $N_{tot}$ is the total number of original unit cells in
the lattice. The quadrupole moment $Q^{\mu\nu}$ becomes finite as long as the system breaks the rotational symmetry, whereas, the dipole moment $P^\mu$ is finite when both
rotational and inversion symmetries are broken. Thus, we denote the phase as nematic phase ($\mathcal{NEM}$) where translation and rotation symmetries are broken but inversion symmetry is preserved satisfying $P^{\mu}\!=\!0,~ Q^{\mu\nu} \! \neq \!0$. On the other hand, we denote the phase with $P^{\mu}\!\neq\!0,~ Q^{\mu\nu}\!\neq\!0$ as ferrielectric phase ($\mathcal{FI}$) where all the symmetries (translation, rotation and inversion) of the system are broken. The magnetic state of the system is characterized
by the arrangement of the magnetization on each sublattices in the magnetic unit cell. Depending
on the magnetic ordering we have ferromagnetic, ferrimagnetic or antiferromagetic phases.

From now on, we highlight possible phases and phase transitions at each integer filling $\nu \!=\![1, 6]$ in the presence of flux per plaquette $\phi \!=\!2\pi/3$, by tuning parameters spin orbit coupling $\lambda_{so}$, onsite interaction $U$ and nearest neighbor interaction $V$. Throughout the paper, the Chern insulator is denoted as `$CI$', normal insulator is denoted as `$NI$' and the translational, rotational and inversion symmetries are denoted as `$T$', `$R$' and `$I$' respectively.

\subsection{$\nu=6$}
For $\nu\!=\!6$, half of the total twelve bands are filled. As shown in Fig.~\ref{fig0}, the non-interacting system is either (semi-) metal or time reversal symmetry broken QSHI depending on the values of $\lambda_{so}$. 
Increase of $U$ in the system stabilizes the Mott insulator with antiferromagnetic ordering. One of the interesting aspect is that the coexistence of SOC and magnetic flux favors 
antiferromagnetic N\'eel order in the $xy$ plane but ferromagnetic order along the $z$ direction, resulting in {\it coplanar} magnetic ordering. In terms of mean-field order parameters, $M^{x(y)}_{\alpha A}\!=\! -M^{x(y)}_{\alpha B} \!=\! M~\forall{\alpha}$ and $M^z_{\alpha a} \!=\!\tilde{M} ~\forall{(\alpha,~ a)}$. 
Such magnetic order could be understood from the perturbation theory at  large $U$ limit. In the presence of SOC, finite magnetic flux  gives rise to the  effective spin-spin interactions which involve three sites on a honeycomb lattice. Eq.~\eqref{eq:perturbation} shows the effective spin Hamiltonian based on the perturbation theory up to the third order in large $U$ limit. 
\begin{widetext}
\begin{eqnarray}
 \mathcal{H}_{eff}&=&\frac{4t^2}{U}\sum_{\langle ij\rangle}\bm S_{i}.\bm S_{j}+\frac{4\lambda_{so}^2}{U}
 \sum_{{\langle\langle ij\rangle\rangle}}(-S^x_iS^x_j-S^y_iS^y_j+S^z_iS^z_j)\nonumber \\
 &\!+\!&
 \frac{t^2\lambda_{so}}{U^2}\sum_{\langle ijk\rangle}
 \sin{\phi_{ijk}}\Bigg[2(S^z_i\!+\!S^z_j\!+\!S^z_k)\!-\!24 S^z_iS^z_jS^z_k 
 \!-\!24\Big(S^z_i(S^x_jS^x_k\!+\!S^y_jS^y_k)\!+\!S^z_k(S^x_iS^x_j\!+\!S^y_iS^y_j)\!-\!S^z_j(S^x_iS^x_k\!+\!S^y_iS^y_k)\Big)\Bigg] 
 \nonumber \\
&\!-\!&\frac{\lambda^3_{so}}{U^2}\sum_{\langle\langle ij k\rangle\rangle}
 \sin{\phi^{'}_{ijk}}\Bigg[2(S^z_i\!+\!S^z_j\!+\!S^z_k)
 -24 S^z_iS^z_jS^z_k\!+\!24\Big(S^z_i(S^x_jS^x_k\!+\!S^y_jS^y_k)\!+\!S^z_k(S^x_iS^x_j\!+\!S^y_iS^y_j)\!+\!S^z_j(S^x_iS^x_k\!+\!S^y_iS^y_k)\Big)\Bigg]\nonumber\\
\label{eq:perturbation}
\end{eqnarray}
\end{widetext}
where $\langle ijk \rangle$ indicates the nearest neighbors sites $i$ and $j$, nearest neighbors sites $j$ and $k$, thus, next nearest neighbors sites $i$ and $k$, and $\langle \langle ijk \rangle \rangle$ indicates $i$,$j$ and $k$ as the second nearest neighbors with each other. $\bm S_i \!=\!(S^x_i,S^y_i,S^z_i)$ is the spin at site $i$ and  $\phi_{ijk}$ is the magnetic flux in the triangle formed
by connecting the $i,j$ and $k$ sites.
Focusing on the third order perturbation terms $\mathcal{O}\Big( t^2 \lambda_{so}/U^2\Big)$ and $\mathcal{O}\Big( \lambda_{so}^3/U^2\Big)$, one can read off that ferromagnetic order along $z$-direction 
with antiferromagnetic N\'eel order on $xy$-plane is preferred. 
With increasing $U$, the contribution
from the second  order correction is  dominant and magnitudes of $S^z$ ordering decrease which is also seen in our numerical results.

The presence of nearest neighbor interactions $V$ induces a charge ordering of the form 
$\Delta_{\alpha A}\!=\!-\Delta_{\alpha B}\!=\!\Delta,~\forall \alpha$. We denote this type of charge ordering as a charge density wave (`$\mathcal{CDW}$'), that breaks only inversion symmetry of the system.
In the  presence of both $U$ and $V$, there is a first order transition from
the $NI$ with coplanar magnetic order  to  the $NI$ with $\mathcal{CDW}$ on increasing  $V/U$.   

\subsection{$\nu \!=\!5$}
\begin{figure}[hbtp]
\begin{center}
\includegraphics[scale=0.4]{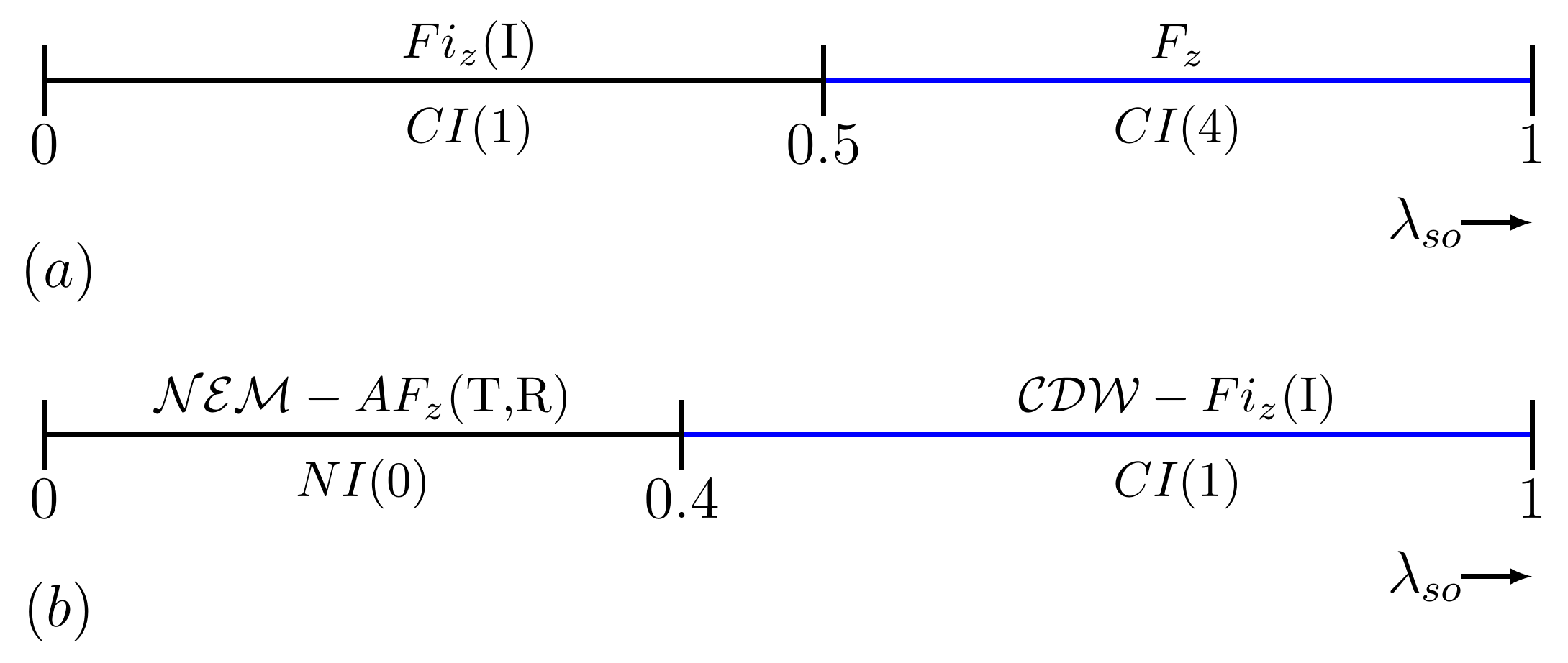}\\
\caption{\label{fig2}(Color online) Phase diagram as a function of $\lambda_{so}$ at filling $\nu=5$ for (a) $U\!=\!4$ and $V\!=\!0$
and (b) $U\!=\!8$ and $V\!=\!2$. 
$F,~Fi$ and $AF$ represents ferromagnetic, ferrimagnetic and antiferromagetic phase respectively. The subscript shows the preferred  magnetization direction. $\mathcal{NEM}$ and $\mathcal{CDW}$ denotes the nematic order and conventional charge density wave respectively. The letter in the bracket shows the type of spatial symmetries broken in the system (translation ($T$), rotation ($R$) and inversion ($I$)).
$CI(\cdot)$ and $NI(\cdot)$ denote Chern insulator and normal insulator respectively, with the total Chern number $(\cdot)$ for occupied bands.}
\end{center}
\end{figure}

For $\nu\!=\!5$, the non-interacting system has several phases as a function of SOC (see Fig.~\ref{fig0}): metal, $CI$ ($C_f \!=\!-2$),  $CI$ ($C_f \!=\!4$). Here, the total Chern number for all filled bands, $C_f$, is always changed by multiple integers of $3$ as a consequence of tripled magnetic unit cell in the presence of flux $2\pi/3$ as long as the translational symmetry is preserved in the system.
The presence of $U$ and $V$ leads to many distinct topological phases accompanied with spontaneous symmetry breaking which will be explained in detail in the next paragraph. Fig.~\ref{fig2} highlights some of those phases as a function of $\lambda_{so}$; (a) $U\!=\!4$ and $V\!=\!0$ and (b) $U\!=\!8$ and $V\!=\!2$. 
In Fig.~\ref{fig2}, the types of charge and magnetic ordering are indicated above each phase diagram line, along with the broken spatial symmetries $T,~R,~I$ listed in the bracket. 
Below the line in Fig.~\ref{fig2}, the type of state, either normal insulator ($NI$) or Chern insulator ($CI$)  along with the total Chern number of the filled bands is shown in the
bracket. 
 
With finite $U$, the system has a phase transition stabilizing a ferromagnetic $CI$ phase ($\bm M_{\alpha, a}\!=\!\bm M,~\forall(\alpha,a)$). The magnetization is along the $z-$ direction for finite SOC and thus the phase is denoted as `$F_zCI$'. The total Chern number remains same as the non-interacting case. On further increasing $U$, there
exist two more phase transitions to (i) ferrimagnetic $CI$ phase (`$Fi_zCI$') with $C_f\!=\!1$ (see Fig.~\ref{fig2}a), the magnetic ordering of the form $M^{z}_{\alpha,A}\!=\!M_A,~M^{z}_{\alpha,B}\!=\!M_B~ \forall \alpha$ where $M_A\!\neq\! M_B$ and are opposite in sign, thus breaking inversion symmetry of the system (ii) ferrimagnetic $NI$ phase (`$Fi_zNI$'),
 $\bm{M}_{\alpha,a}^z$ is different for all $\alpha,a$ breaking the entire symmetries of the lattice. In the later case, we can notice that the difference in the Chern number is no longer multiple integers of $3$ as the translational symmetry of the system is broken. The critical value of $U$ for these phase transitions increases with the increase in the spin
 orbit coupling strength.

The nearest neighbor interaction $V$ induces charge ordering similar to that of $\nu\!=\!6$ case and the system becomes a Chern insulator with $\mathcal{CDW}$, labeled as $`\mathcal{CDW}CI'$ with $C_f\!=\!1$. In the limit $V \!\ll\! U$ and small SOC, the system is a trivial insulator with 
$ M_{1A}^z\!=\!- M^z_{3B},~M^z_{2A}\!=\!- M^z_{2B},~ M^z_{3A} \!=\!- M^z_{1B}$ where $z$ direction magnetization is preferred by SOC and 
$\Delta_{1A} \!=\! \Delta_{3B},~\Delta_{2A}\!=\!\Delta_{2B},~\Delta_{3A}\!=\! \Delta_{1B}$.
Here, rotation and translation symmetries are broken but inversion symmetry is preserved in the system.
In terms of charge order, the dipole moment $P^\mu$ is zero but the quadrupole moment $Q^{\mu \nu}$ is finite, as expected on symmetry argument and the system is in the nematic phase. The phase with this type of magnetic and charge ordering is referred as the nematic-antiferromagnetic (along $z$ direction) normal insulator, `$\mathcal{NEM}\!-\!AF_zNI$' as listed in Fig.~\ref{fig2}b.
For larger SOC, the system goes into a $\mathcal{CDW}$ phase with staggered magnetization breaking
the inversion symmetry of the system but preserving all other symmetries.
The magnetic ordering is of the form  $M^{z}_{\alpha,A}\!=\!M_A,~M^{z}_{\alpha,B}\!=\!M_B~ \forall \alpha$.
Here, $M_A\!\neq \!M_B$ and are opposite in sign. This is the $\mathcal{CDW}$ ferrimagnetic Chern insulator with $C_f\!=\!1$ denoted as 
`$\mathcal{CDW}\!-\!Fi_zCI$' shown in Fig.\ref{fig2}b.

\subsection{$\nu=4$}
\begin{figure}[hbtp]
\begin{center}
\includegraphics[scale=0.4]{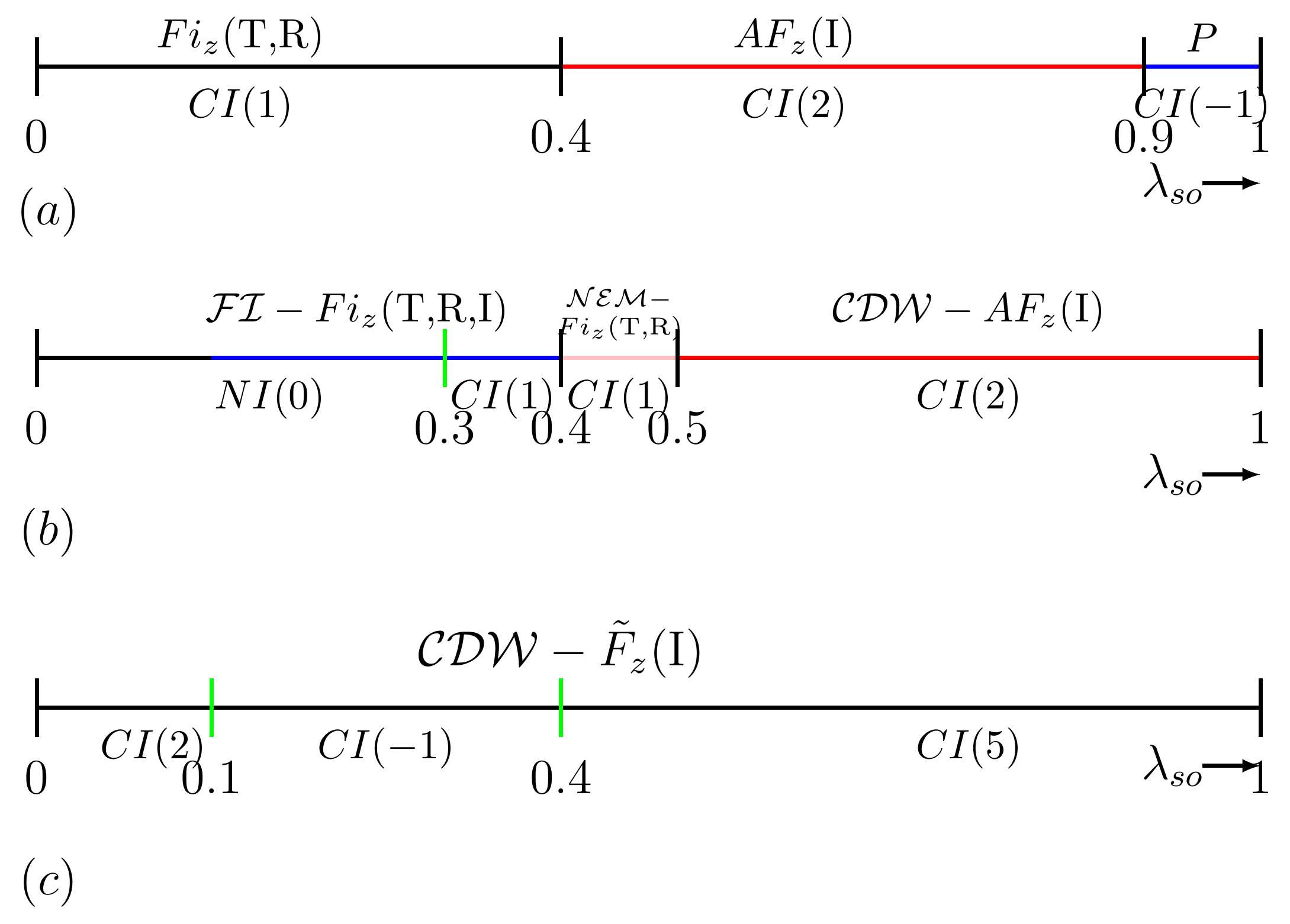}\\
\caption{\label{fig3}(Color online) Phase diagram as a function of $\lambda_{so}$ for (a) $U\!=\!6$ and $V\!=\!0$, 
(b) $U\!=\!6$ and $V\!=\!2$ and (b) $U\!=\!6$ and $V\!=\!8$ at filling $\nu\!=\!4$. $P,~AF,~Fi$ and $~\tilde F$ represents paramagnetic, antiferromagnetic,  ferrimagnetic and ferro- but staggered magnetization respectively. $\mathcal{CDW},~\mathcal{NEM}$ and $\mathcal{FI}$ denotes conventional charge density wave, nematic and  ferrielectric phase respectively.
The letter in the bracket shows the type of spatial symmetries broken in the system (translation ($T$), rotation ($R$) and inversion ($I$)).
$CI(\cdot)$ and $NI(\cdot)$ denote Chern insulator and normal insulator respectively, with the total Chern number $(\cdot)$ for occupied bands.}
\end{center}
\end{figure}

At $\nu\!=\!4$, the non-interacting system realizes three phase transitions between different Chern insulators ($CI$) with $C_f\!=\! 2,-1$ and $5$ (See Fig.~\ref{fig0}).
For different values of $U$ and $V$, Fig.~\ref{fig3} highlights the possible normal/ Chern insulating phases with charge order and magnetic order at (a) $U\!=\!6$ and $V\!=\!0$, (b) $U\!=\!6$ and $V\!=\!2$ and (c) $U\!=\!6$ and $V\!=\!8$.

In the presence of only $U$ and small SOC, there is a phase transition from paramagnetic $CI$ phase (`$PCI$') to $CI$ phase with 
 ferrimagnetic ordering along $z$ direction (`$Fi_zCI$'). In $Fi_zCI$ phase, the magnetic order parameters are the form of {${M^z}_{A1}\!=\!{M^z}_{B3},~{M^z}_{A2}\!=\!{M^z}_{B2},~{M^z}_{A3}\!=\!{M^z}_{B1}$} (See Fig.~\ref{fig3}a). By stabilizing this type of magnetic order, the system breaks translation and rotation symmetries but preserves the inversion symmetry. For intermediate to larger values of
$\lambda_{so}$, increase of $U$ leads to the transition from $PCI$ to antiferromagetic Chern insulator, $AF_zCI$  ($M^z_{\alpha,A}\!=\!-M^z_{\alpha,B}\!=\!M,~ \forall \alpha$) which breaks only the inversion symmetry of the system (see Fig.~\ref{fig3}a). 

In the presence of $V$, for $V\! \ll \!U$, the system can develop different types of magnetic order and charge order depending on the SOC strength: (i) 
${M^z}_{A1}\!=\!{M^z}_{B3},~{M^z}_{A2}\!=\!{M^z}_{B2},~{M^z}_{A3}\!=\!{M^z}_{B1}$ and ${\Delta}_{A1}\!=\!{\Delta}_{B3},~{\Delta}_{A2}\!=\!{\Delta}_{B2},~{\Delta}_{A3}\!=\!{\Delta}_{B1}$ which preserves 
only inversion symmetry of the system (ii) $M_{\alpha,a}$ and $\Delta_{\alpha,a}$ is different for all sublattices which breaks all the symmetries of the system. While the former phase (i) corresponds to the nematic-ferrimagnetic phase ($\mathcal{NEM}-Fi_z$), the later one (ii) is the ferrielectric-ferrimagnetic phase ($\mathcal{FI}-Fi_z$). These phases can be either $NI$ or $CI$ ($C_f\!=\!1$) depending on $\lambda_{so}$ for a particular values $U$ and $V$ (see Fig.~\ref{fig3}b). For large SOC, the system goes into the inversion symmetry broken phases described by order parameters $\Delta_{\alpha,A}\!=\!-\Delta_{\alpha,B}\!=\!\Delta,~ \forall \alpha$, $M^z_{\alpha,A}\!=\!-M^z_{\alpha,B}\!=\!M,~ \forall \alpha$. This is a $\mathcal{CDW}-AF_zCI$ phase with $C_f=2$ as shown in Fig.~\ref{fig3}b. For large $V$ limit, the system goes into the phase with charge density wave and ferro- but staggered magnetic order phase, (`$\mathcal{CDW}-\tilde F_zCI$'). In details, magnetic and charge ordering is of the form $M^z_{\alpha,A}\!=\!M_A,~M^z_{\alpha,B}\!=\!M_B,~\forall \alpha$ and $M_A\neq M_B$ but have same sign and 
$\Delta_{\alpha,A}=\Delta=-\Delta_{\alpha,B},~ \forall \alpha$. In this phase, the change of Chern numbers is similar to the non-interacting case with increasing SOC. (See Fig.~\ref{fig3}c). Indeed, the Chern numbers are changed by multiple integers of 3 here due to the presence of translational symmetry.

\subsection{$\nu=3$}
\begin{figure}[hbtp]
\begin{center}
\includegraphics[scale=0.38]{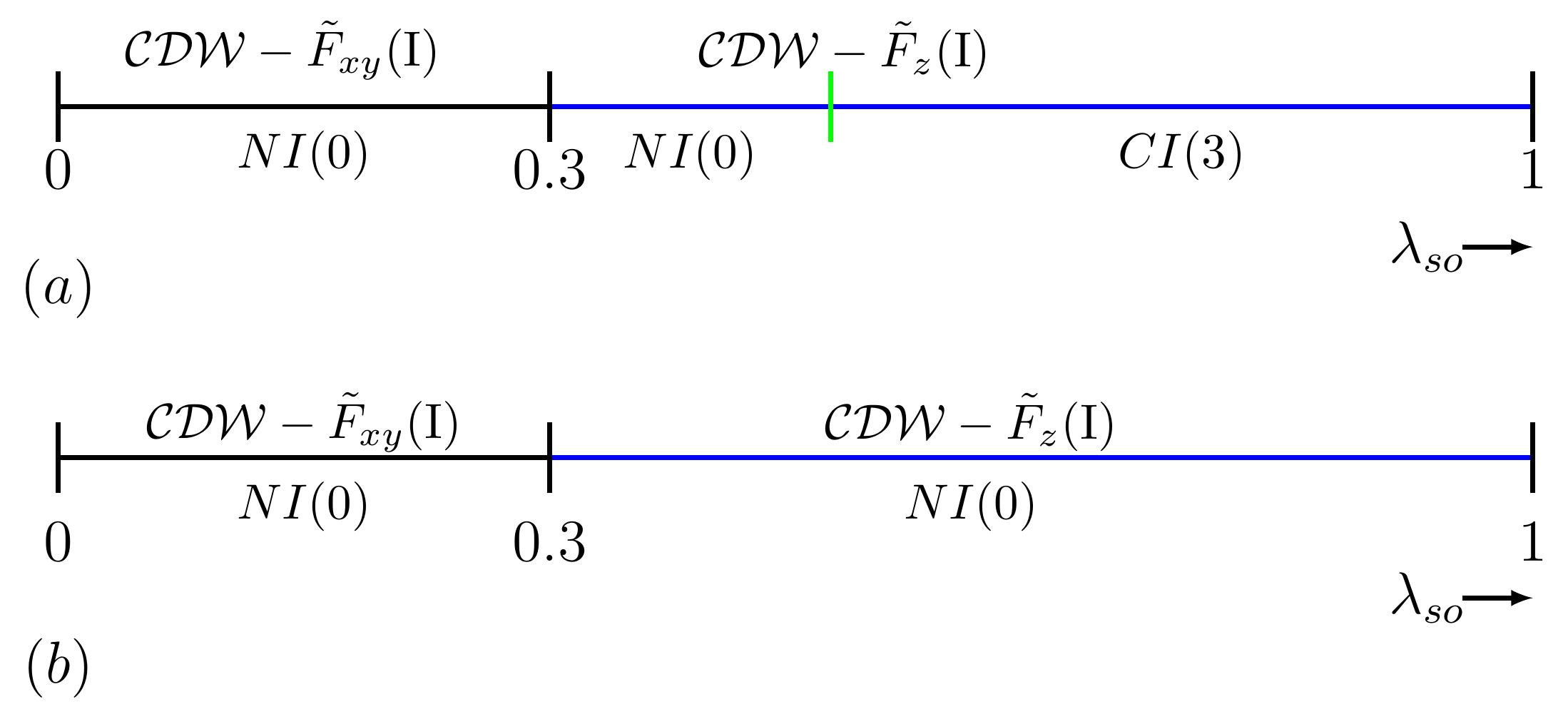}\\
\caption{\label{fig4}(Color online) Phase diagram as a function of $\lambda_{so}$ for (a) $U\!=\!6$ and $V\!=\!4$
and (b) $U\!=\!6$ and $V\!=\!8$ at filling $\nu\!=\!3$. $\mathcal{CDW}-\tilde F$ is the charge density wave with ferro- but staggered magnetization. The subscript denotes the magnetization direction.
The letter in the bracket shows the type of spatial symmetries broken in the system (translation ($T$), rotation ($R$) and inversion ($I$)).
$CI(\cdot)$ and $NI(\cdot)$ denote Chern insulator and normal insulator respectively, with the total Chern number $(\cdot)$ for occupied bands.}
\end{center}
\end{figure}

At filling $\nu \!=\!3$, the  system   is  in a metallic  phase similar to $\nu\!=\!5$ in the absence of SOC and interactions. In the presence of SOC, a  gap opens up at $\lambda_{so}=0.1$  and the system goes into a QSHI phase (See. Fig.~\ref{fig0}). 
There exists a band crossing at $\lambda_{so}=0.41$
and on further increasing the SOC the system goes into a $CI$ with $C_f\!=\!3$ (see Fig.~\ref{fig0}).
With finite $U$ and $V$, the system develops charge density wave ($\mathcal{CDW}$) and ferro- but staggered magnetization along either $xy \text{ or } z$ direction ($\tilde{F}_z$ or $\tilde{F}_{xy}$). While phase with magnetic ordering $\tilde{F}_z$ can be $CI$ or $NI$ depending on SOC, phase with magnetic ordering $\tilde{F}_{xy}$ is always $NI$. These phases only break inversion symmetry and are labeled as either `$\mathcal{CDW}-\tilde F_\mu NI$' or `$\mathcal{CDW}-\tilde F_z CI$' in Fig.~\ref{fig4}, where $\mu$ represents $xy$ or $z$ direction. Fig.~\ref{fig4} gives the line phase diagram as a function of $\lambda_{so}$ for (a) $U=6,~V=4$ and (b) $U=6,~V=8$.
The magnetization direction changes from $xy$ to $z$ on increasing $\lambda_{so}$. It is worth to mention that in the absence of magnetic flux, the same extended Kane-Mele-Hubbard model has already been explored and similar phase transitions are discussed in Ref.~\onlinecite{mishra2018}.

\subsection{$\nu=2$}
\begin{figure}[hbtp]
\begin{center}
\includegraphics[scale=0.4]{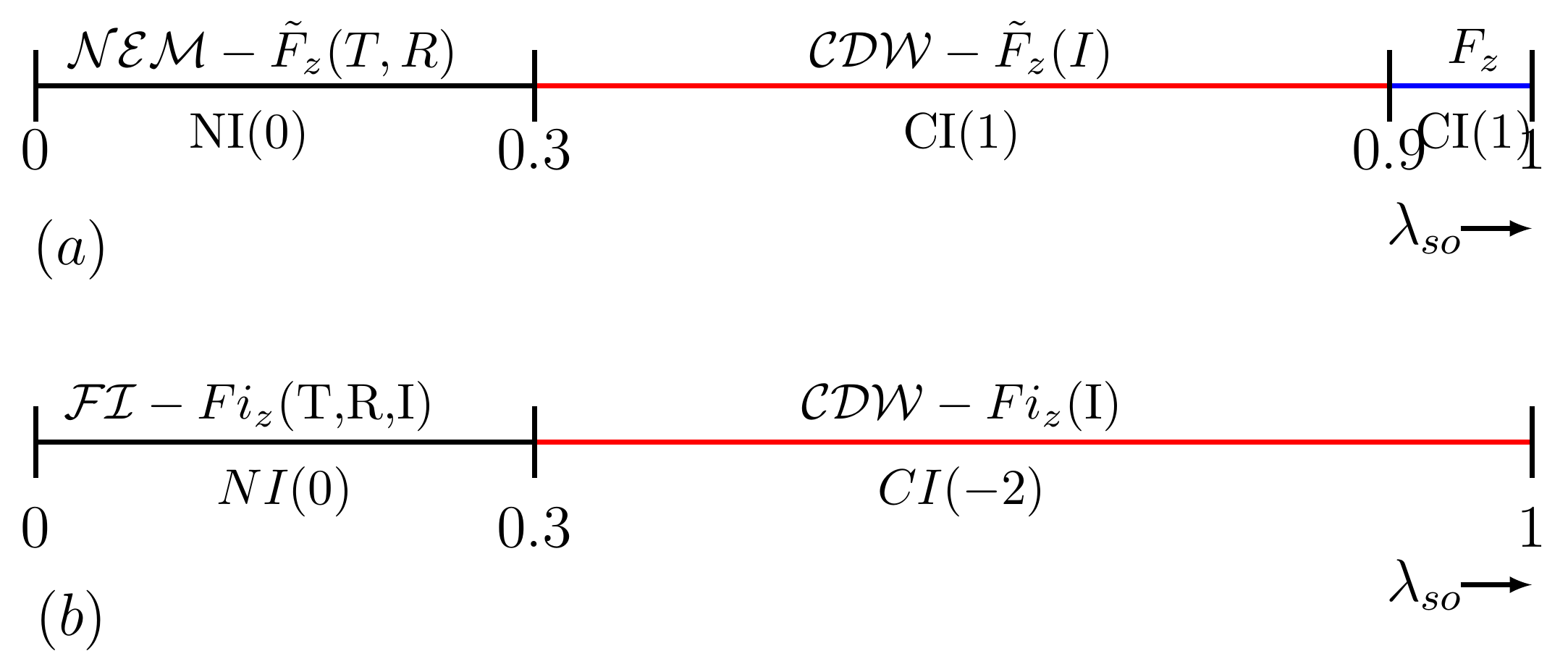}\\
\caption{\label{fig5}(Color online)The phase diagram as a function of $\lambda_{so}$ for (a) $U\!=\!8$ and $V\!=\!4$
and (b) $U\!=\!2$ and $V\!=\!8$ at filling $\nu\!=\!2$. $F,~\tilde F~ \mbox{and}~ Fi$ denotes ferromagnetic, ferro- but staggered magnetization and ferrimagnetic phase respectively. $\mathcal{NEM}$ and $\mathcal{CDW}$ represents Nematic phase and conventional charge density wave phase respectively.
The letter in the bracket shows the type of spatial symmetries broken in the system (translation ($T$), rotation ($R$) and inversion ($I$)).
$CI(\cdot)$ and $NI(\cdot)$ denote Chern insulator and normal insulator respectively, with the total Chern number $(\cdot)$ for occupied bands.}
\end{center}
\end{figure}

At $\nu=2$, the non-interacting system favors either the $CI$ phase with $C_f\!=\!-2$ for $\lambda_{so} \!\leq \!0.37$ or the metallic phase for other values of SOC. 
In the presence of both $U$ and $V$, Fig.~\ref{fig5} represents the possible phases as a function of $\lambda_{so}$ especially for (a) $U\!=\!8$ and $V\!=\!4$, (b) $U\!=\!2$ and $V\!=\!8$.
With increasing $U$ only, the system develops the ferromagnetic ($M^z_{\alpha,a}\!=\!M^z,~ \forall(\alpha,a)$) $CI$ phase (`$F_zCI$') from paramagnetic phase. In this case, the spin up and down bands are separate and there are two bands with the same spin below the Fermi level. The Chern numbers of these two lowest bands are $-1$ and $2$ which results in the total Chern number of the filled bands to be $C_f=1$.
Including small $V$, 
  at intermediate to large SOC, there is a phase transition to the $CI$ phase ($C_f\!=\!1$) with charge density wave and ferromagnetic order along $z$ direction (`$\mathcal{CDW}-\tilde {F}_zCI$') phase as shown in Fig.~\ref{fig5}a. For small SOC, the charge and magnetic ordering is of 
the form $\Delta_{A1}\!=\!\Delta_{B3},~\Delta_{A2}\!=\!\Delta_{B2},~\Delta_{A3}\!=\!\Delta_{B1}$ and 
$M^z_{A1}\!=\!M^z_{B3},~M^z_{A2}\!=\!M^z_{B2},~M^z_{A3}\!=\!M^z_{B1}$. Thus, the system  is a normal insulator with nematic and ferromagnetic order, (`$\mathcal{NEM}-\tilde {F}_zNI$'), as seen in Fig.~\ref{fig5}a. For $V\!>\!U$ and intermediate to high SOC, the system is in the $CI$ phase ($C_f\!=\!-2$) with charge density wave and ferrimagnetic along $z$ direction ordering (`$\mathcal{CDW}-Fi_zCI$') as shown in Fig.~\ref{fig5}b.   
In this case, the filled bands consist of spin up and down bands each with Chern number $-1$
giving $C_f=-2$. For very small SOC, the magnetic and charge ordering are different for all sublattices, resulting in the $\mathcal{FI}-Fi_zNI$ phase.


\subsection{$\nu=1$}
\begin{figure}[hbtp]
\begin{center}
\includegraphics[scale=0.4]{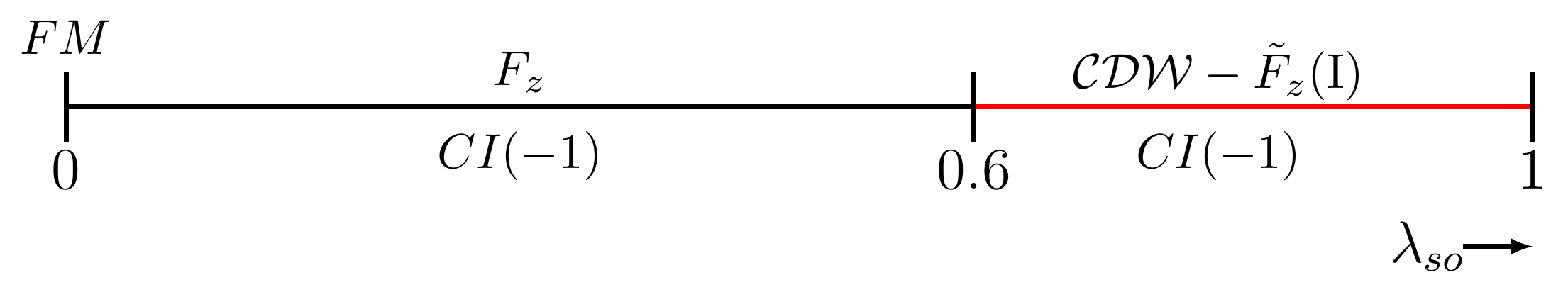}\\
\caption{\label{fig6}(Color online)The phase diagram as a function of $\lambda_{so}$ for $U\!=\!3$ and $V\!=\!6$ at filling $\nu\!=\!1$.
$F$ represents ferromagnetic phase and $\mathcal{CDW}-\tilde F$ represents charge density wave with ferro- but staggered magnetization.
The letter in the bracket shows the type of spatial symmetries broken in the system (translation ($T$), rotation ($R$) and inversion ($I$)).
$CI(\cdot)$ and $NI(\cdot)$ denote Chern insulator and normal insulator respectively, with the total Chern number $(\cdot)$ for occupied bands.}
\end{center}
\end{figure}

When there is a single electron in the entire magnetic unit cell, the non interacting system exhibits either metallic phase or $CI$ with $C_f\!=\!-1$ shown in Fig.~\ref{fig0}.  
Fig.~\ref{fig6} represents the phase diagram as a function of $\lambda_{so}$ for $U\!=\!3$ and $V\!=\!6$.
In the presence of both $U$ and $V$, as shown in Fig.~\ref{fig6}, the system develops either the $CI$ phase with pure ferromagnetic order along $z$ direction (`$F_zCI$') or the $CI$ phase with both charge density wave and ferro- but staggered magnetization along $z$ direction (`$\mathcal{CDW}-\tilde F_zCI$'). The Chern number $C_f$ remains unchanged under these transitions, thus $C_f\!=\!-1$. 

\subsection{Staggered flux and Orbital currents}
\begin{figure}[h!]
\begin{center}
\includegraphics[scale=0.12]{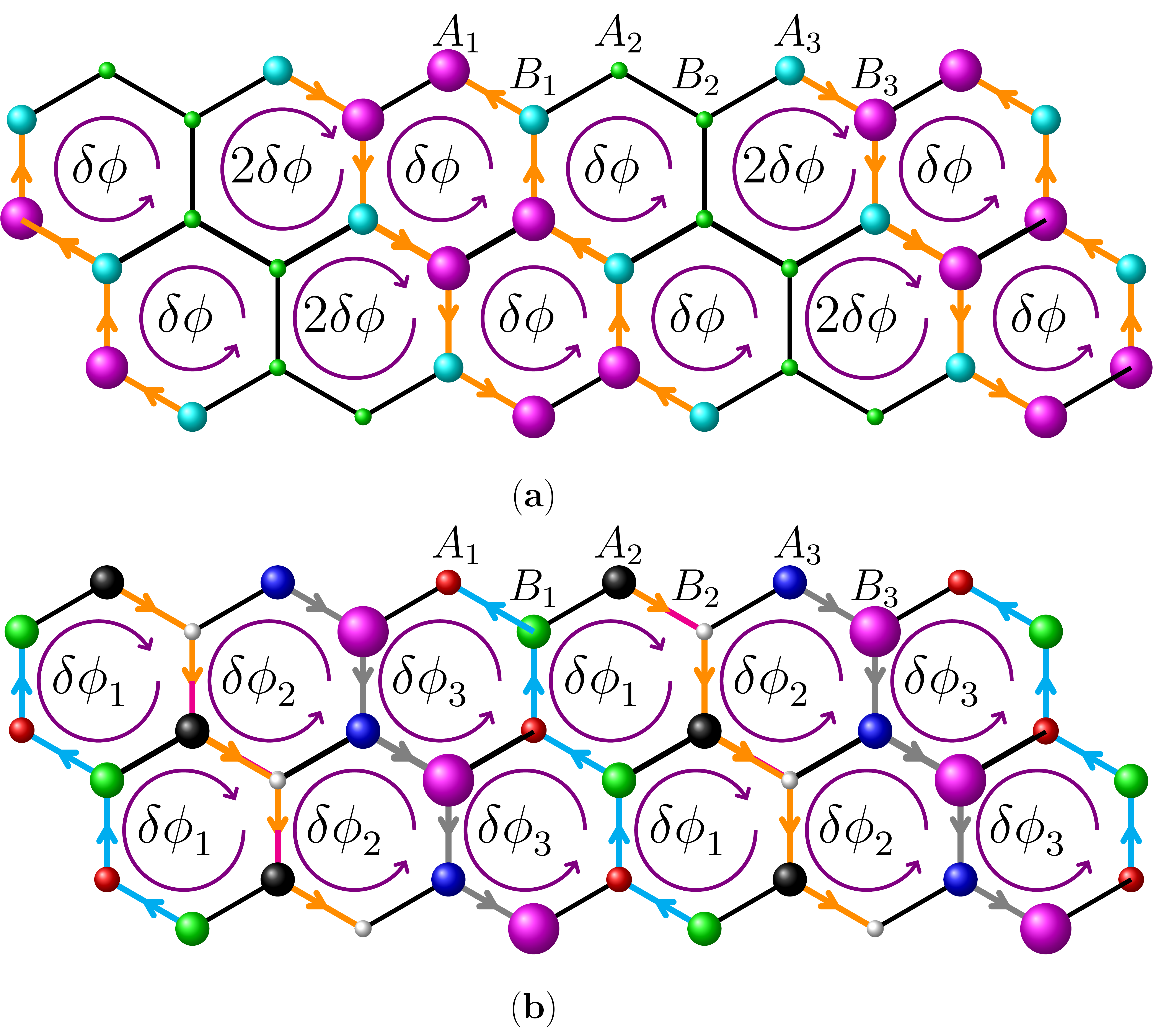}
\caption{\label{currents}(Color online) Schematic picture of staggered flux in hexagonal plaquette and
  current flow on the links for (a) nematic
  phase (b) ferrielectric phase. The flux distribution is shown after
  subtracting out the background magnetic flux $2\pi/3$. In ferrielectric
  phase, $\delta\phi_1\!=\!\delta\phi_3+\delta\phi_2$.  The bonds with different colored 
  arrows are the links with distinct magnitudes of finite current. }
\end{center}
\end{figure}

For particular integer fillings $\nu\!=\!4$ and 5, there are phases with broken translation and rotation symmetries. (See the phases marked with ($T,R$) or ($T,R,I$) in Fig.~\ref{fig2} and Fig.~\ref{fig3}). These types of phases are always accompanied with finite anisotropic bond order parameters $\chi_{(\alpha,a),(\beta,b)}^{\sigma \sigma}$ and result in staggered flux and non-zero currents on the links. 
Fig.~\ref{currents} shows the examples of 
charge order, link current and flux distribution in every hexagonal plaquettes on the lattice. 
Fig.~\ref{currents}a represents the case of nematic ($\mathcal{NEM}$) phase where the current
on the $A_1B_1$ and $A_3B_3$ links have the same magnitude but are in opposite directions and current on $A_2B_2$ is zero
such that the inversion symmetry of the system is preserved. 
Fig.~\ref{currents}b represents the case of ferrielectric ($\mathcal{FI}$) phase where the current on the links
$A_1B_1$, $A_2B_2$ and $A_3B_3$ are different, thus, breaking the inversion symmetry of the system. 

\section{Conclusion}
In conclusion, we have studied the extended Kane-Mele-Hubbard model in the Hofstadter regime on the honeycomb lattice. In particular, focusing on the magnetic flux $2\pi/3$ and considering onsite and nearest-neighbor interactions, we showed the existence of various complex phases due to the interplay of  electron interactions
and SOC in the Hofstadter regime. Within the mean field analysis, the different types of normal or Chern insulating phases accompanied with charge order and magnetic order have been theoretically proposed. In particular, at filling $\nu\!=\!6$ where half of the total 12 bands are occupied, the normal insulator with coplanar magnetic order is uniquely  stabilized due to the combination of spin orbit coupling and magnetic flux. At other integer fillings especially at $\nu\!=\!4$ and 5, there exist the Chern insulating phases where both electric dipole (quadrupole) moments and ferrimagnetic moments coexist. Such topological {\it multiferroic} phases have interesting aspects like staggered magnetic flux and orbital currents in addition to finite Hall conductivity. 

 The recent experimental
observation of Hofstadter butterfly in graphene superlattice \cite{dean2013,ponomarenko2013,hunt2013,yu2014} 
and the realization of the Hofstadter Hamiltonian in the optical lattice \cite{aidelsburger2013,miyake2013} 
have opened a new arena to study the effect of interaction in this strong
magnetic flux regime. 
The superlattice of transition metal chalcogenides series could be the potential candidates to explore such exotic phases we propose here, where magnetic field, SOC and electron interactions could play a significant role. Moreover, in ultra cold atomic system the ability to tune the control parameters over wide ranges, makes it an ideal place
to study. 
While magnetic field and SOC can be implemented artificially in the optical lattice by introducing synthetic gauge fields \cite{jaksch2003,gerbier2010,aidelsburger2011,aidelsburger2013,aidelsburger2013a,struck2012,jotzu2014,goldman2014,
goldman2010,cooper2011,hauke2012,mazza2012,kennedy2013,beeler2013,struck2014,gorecka2011}, interaction can be tuned
in the lattice by Feshbach resonances \cite{bloch2008}. 
Furthermore, the non-trivial topology of the band can also be determined in the optical lattice system \cite{alba2011,zhao2011,price2012,sriluckshmy2014,aidelsburger2015,liu2010,buchhold2012,goldman2012a}.
Hence, both superlattice of 2D van-der Waals materials and optical lattices enable us to investigate new topological {\it multiferroic} phases with experimental realization in future.


\noindent
{ Acknowledgments---}
The authors acknowledge support from BK21plus and National Research Foundation Grant (NRF-2017R1A2B4008097). Hospitality
at APCTP during the program ``Asia Pacific Workshop on Quantum Magnetism" is kindly acknowledged. S.B.L thanks
the hospitality at the Physics Department of University of
California, San Diego. 



\end{document}